\documentclass[conference]{IEEEtran}
\pdfoutput=1 
\makeatletter
\def\ps@headings{%
\def\@oddhead{\mbox{}\scriptsize\rightmark \hfil \thepage}%
\def\@evenhead{\scriptsize\thepage \hfil \leftmark\mbox{}}%
\def\@oddfoot{}%
\def\@evenfoot{}}
\makeatother
\usepackage{tabularx}
\usepackage{amssymb}
\usepackage{graphicx}
\usepackage{tikz}
\usepackage{amsmath}
\usepackage{mathtools}
\usepackage{subcaption}
\usepackage{pgfplots}
\usepackage{color, colortbl}
\usepackage{xcolor}
\usepackage[utf8]{inputenc}
\usepackage{tabularx,ragged2e,booktabs}
\usepackage{makecell}
\usetikzlibrary{matrix,positioning}

\usepgfplotslibrary{groupplots}

\newcolumntype{C}{>{\Centering\arraybackslash}X} % centered "X" column
\definecolor{Gray}{gray}{0.9}
\pgfplotsset{width=0.8\linewidth,compat=1.14,
	    every axis/.append style={
		label style={font=\small},
		tick label style={font=\small},
		enlarge x limits={abs=0.25}},
 		 /pgfplots/ybar legend/.style={
		/pgfplots/legend image code/.code={%
		\draw[##1,/tikz/.cd,yshift=-0.25em]
		(0cm,0cm) rectangle (3pt,0.8em);},
},
}
\IEEEoverridecommandlockouts
\IEEEpubid{\makebox[\columnwidth]{Version: May 1, 2019\hfill} \hspace{\columnsep}\makebox[\columnwidth]{}} 
\begin{document}
	
\title{DSAF: Dynamic Slice Allocation Framework for 5G Core Network}
\author{\IEEEauthorblockN{Danish Sattar and Ashraf Matrawy\\}
	\IEEEauthorblockA{
		Carleton University,
		Ottawa, ON Canada\\
		Email: \{Danish.Sattar, Ashraf.Matrawy\}@carleton.ca}}
\maketitle

\begin{abstract}
Network slicing is a key to supporting different quality-of-service requirements for users and application in the 5G network. However, allocating network slices efficiently while providing a minimum guaranteed level of service in a mobile core is challenging. To address this question, in our previous work we proposed an optimization model to allocate slices. It provided a static and manual allocation of slices. In this paper, we extend our work to dynamically allocated slices. We propose a dynamic slice allocation framework for the 5G core network. The proposed framework provides user-interaction to request slices and any required services that need to run on a slice(s). It can accept a single or multiple allocation requests, and it dynamically allocates them. Additionally, the framework allocates slices in a balanced fashion across available resources. We compare our framework with the First Come First Serve and First Available allocation scheme. 
\end{abstract}

\IEEEpeerreviewmaketitle

\section{Introduction}
Current mobile networks are static and highly centralized. For instance, 4G core networks are composed of monolithic purpose-built network elements placed in a few data centers \cite{8116371}. With ever-growing data volume, elastic demand for resources and agile application deployment cycle, it has been challenging to meet these demands in the current mobile network architecture. It is expected that by 2020, there will be a 1000-fold increase in data volume and 100-fold in connected devices~\cite{ICT-31766}.
The $5^{th}$ generation (5G) of the mobile network has been proposed to overcome the shortcomings of the current mobile networks and meet future demands. The 5G network architecture relies on Network Slicing to enable agile and rapid development of applications. Network slicing exploits the concept of Network Function Virtualization (NFV) to split a physical infrastructure into multiple virtual networks \cite{8116371} and distribute them in an on-demand fashion. 
In 5G networks, an end-to-end network slice is defined as a complete logical network that includes Radio Access Network (RAN) and Core Network (CN)~\cite{rfcns}. However, it is possible to instantiate RAN and CN slices independently. Network slicing in a 5G network presents a unique challenge that is not present in the previous or current mobile networks. The challenge is how to allocate slices optimally and dynamically to efficiently use the mobile network resources as well as guarantee minimum requested resources. 

To address this question, we proposed an optimization model for allocating slices in the 5G core network \cite{8676260}. The proposed model was designed to allocate slices based on Central Processing Unit (CPU) utilization and link delay. The optimization model provided intra-slice isolation for network functions within a slices. Intra-slice isolation provides a variable degree of physical separation between slice components. For instance, if a slice requires an intra-slice isolation level that is equal to 1, then only a single component (network function) of the slice will be hosted on a given hypervisor. The allocation was \textit{mathematically simulated} through MATLAB. All slice requests were provided as one input to the optimization model. This work was extended for utilization in DDoS mitigation, and it was evaluated by using a real testbed \cite{secureslicingCNS2019}. However, the allocation of the slices in the tested was done \textit{manually} and in a \textit{static} fashion and the slice requests were still provided as one input to the optimization model. There was no mechanism to deal with requests arriving in real-time.  

We address these points in this paper, where we propose a \textbf{D}ynamic \textbf{S}lice \textbf{A}llocation \textbf{F}ramework (DSAF). Our \textbf{contributions} are (1) DSAF can allocate slice \textit{dynamically} in a real-time (2) it can calculate allocation solution (where feasible) for individual or multiple slice requests and update network topology accordingly, and (3) the proposed framework allocates slices in a balanced fashion (i.e., it spreads them across the available resources). DSAF can perform seamless slice allocation and deallocation as well as provide on-demand intra-slice isolation. Only user interaction required in DSAF is when a user requests a slice allocation; every other procedure is automated. DSAF implements our optimization model that can fulfill several requirements of the 5G mobile core network. We compared our proposed framework with the First Come First Serve and First Available (FCFSFA) allocation scheme. Both are evaluated on a real testbed. 

The rest of this paper is organized as follows. In Section~\ref{sec:DSAF:relatedwork}, we present the literature review on 5G network slicing and testbeds. Section~\ref{sec:DSAF:mathmodel} provides an overview of the optimization model for 5G network slicing. In section \ref{sec:DSAF:DSAF}, we present our Dynamic Slice Allocation Framework. We discuss our experiment setup in section \ref{sec:DSAF:expsetup}. In the section~\ref{sec:DSAF:resutls}, we discuss our results and lastly, section~\ref{sec:DSAF:con}, we present our conclusion.
 \begin{figure*}[!ht]
	\centering
	\includegraphics[width=\linewidth,keepaspectratio=true]{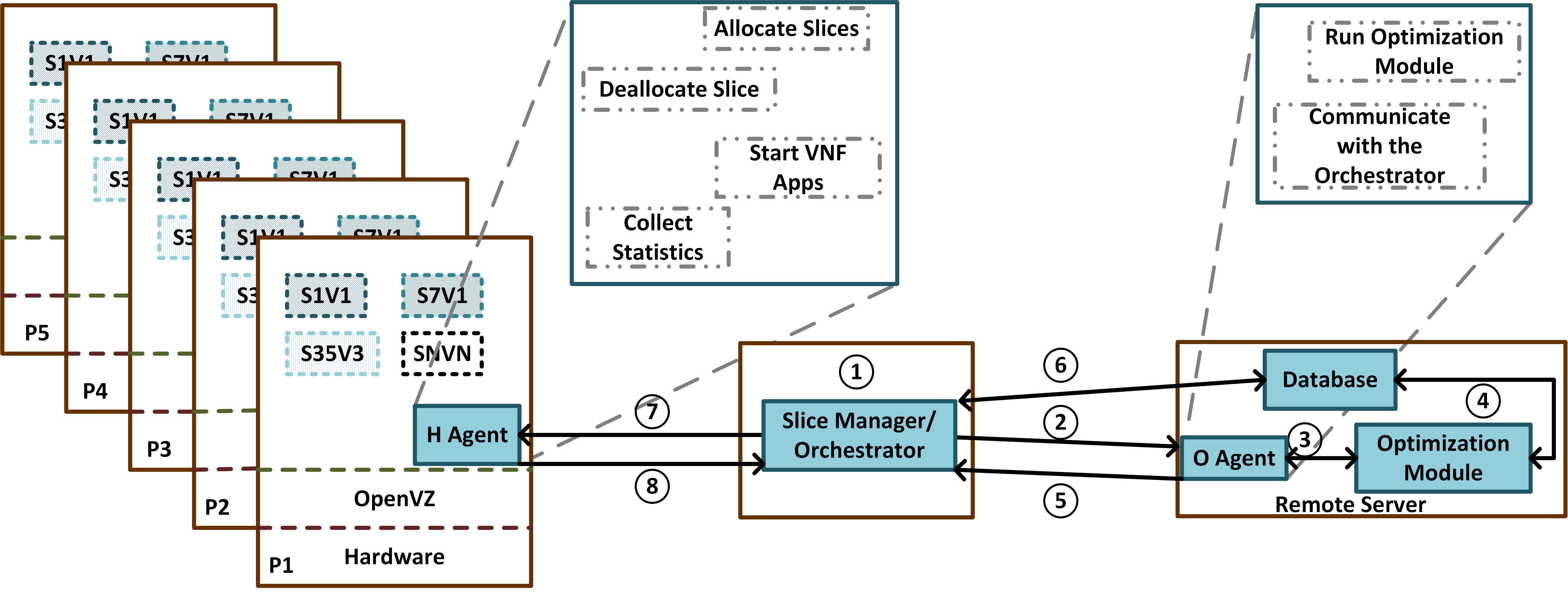}
		\caption{DSAF Logical Topology in our experiments: The brown boxes represent the physical servers and blue boxes represent the framework components. Solid lines show the logical communication paths between framework components}
	\label{fig:SDNTopology}
\end{figure*}
\section{Related Work}
\label{sec:DSAF:relatedwork}
In this section, we discuss existing works on 5G network slice allocation and testbeds. 

\textbf{Slice Allocation:}
M. Jiang \textit{et al.} \cite{7499297} proposed a novel heuristic-based admission control mechanism to allocate slices dynamically. In the proposed scheme, two-tier priority levels are used to allocate slices. In order to maximize the quality of experience for the users inside the slices, effective throughput is measured, and service is provided based on the inter-slice priority. The proposed scheme can dynamically change allocated resources to accommodate higher priority slices.
X. Zhou \textit{et al.} \cite{7509393} discussed providing Network Slice as a Service (NSaaS).  It can be categorized into three classes, i.e., business to business, business to consumer and business to business to consumer. NSaaS could also have three scenarios, i.e., industrial slice, monopolized slice, and event slice. Authors discussed the network slice creation and management as well as mapping of functions, service level agreements and different vendors. An auction-based model for network slicing in 5G has been proposed by M. Jiang \textit{et al.} \cite{7996490}. The objective of the auction-based model is to allocate slices to maximize network revenue. The model takes into consideration the demand and provisions in the network to decide on the price of the network slice. A digital market place (Network Store) for network applications and network functions has been proposed by N. Nikaein \textit{et al.} \cite{Nikaein:2015:NSE:2795381.2795390}. The proposed Network Store would act similar to a Play store for Android or App store for iOS applications. It will offer hundreds of services that would be available to every slice. Authors demonstrated the proposed Network Store on a real testbed and evaluated the performance. The testbed used LTE as radio, open-air EPC to emulate core network and OpenStack to provide virtualization. A. Baumgartner \textit{et al.} \cite{7116162} discussed virtual network function placement in the mobile network core. They proposed a mathematical model that takes into consideration physical mobile core network resources and finds the optimal mapping for requested virtual network functions. The optimization model was evaluated using simulations.

\textbf{5G Testbeds:}
A Practical Open Source Solution for End-to-End Network Slicing (POSENS) has been proposed by G. Aviles \textit{et al.} \cite{8524891}. POSENS uses open source software and hardware to create end-to-end slices. Authors used srsLTE for radio access network and open-air interface EPC for the core network. The independence of slices and performance throughput is discussed in the paper. It also supports an efficient and flexible deployment of network slices. L. Zanzi \textit{et al.} \cite{8491249} demonstrated a real testbed named OVerbooking NEtwork Slices (OVNES). In the testbed, authors considered three different vertical slices, i.e., Public Safety communications, enhanced Mobile BroadBand (eMBB) for voice calls, and eMBB for Internet (best-effort). OpenEPC is used to emulate mobile core and several LTE devices to generate traffic. 

\section{Optimization Model}
\label{sec:DSAF:mathmodel}
In our previous work \cite{secureslicingCNS2019}, we proposed an optimization model to mitigate DDoS attacks. The proposed model mitigated DDoS attacks using intra-slice (between slice components) and inter-slice\footnote{Please note that in this paper we did not consider the inter-slice isolation} (between slices) isolation. In addition to providing defense against DDoS, it can optimally allocate slices. Our model allocated slices to the least utilized servers and finds the minimum delay path. The optimization model also fulfills several requirements of the 5G network. It can guarantee the end-to-end delay and provide different levels of slice isolation for reliability and availability as well as it assures that allocation does not exceed the available system resources. In our model, we only considered CPU, bandwidth, VNF processing delay, and link delay. Intra-Slice isolation could increase the availability of a slice. If all components of the slice are hosted on the same hypervisor, any malfunction could result in the slice unavailability. However, different levels of intra-slice isolation can ensure that full or partial slice remains available. More details of the optimization model can be found in \cite{secureslicingCNS2019}. We use this model as the optimization model in this paper.
\begin{figure}[!ht]
	\centering
	\includegraphics[height=0.8\linewidth,keepaspectratio=true]{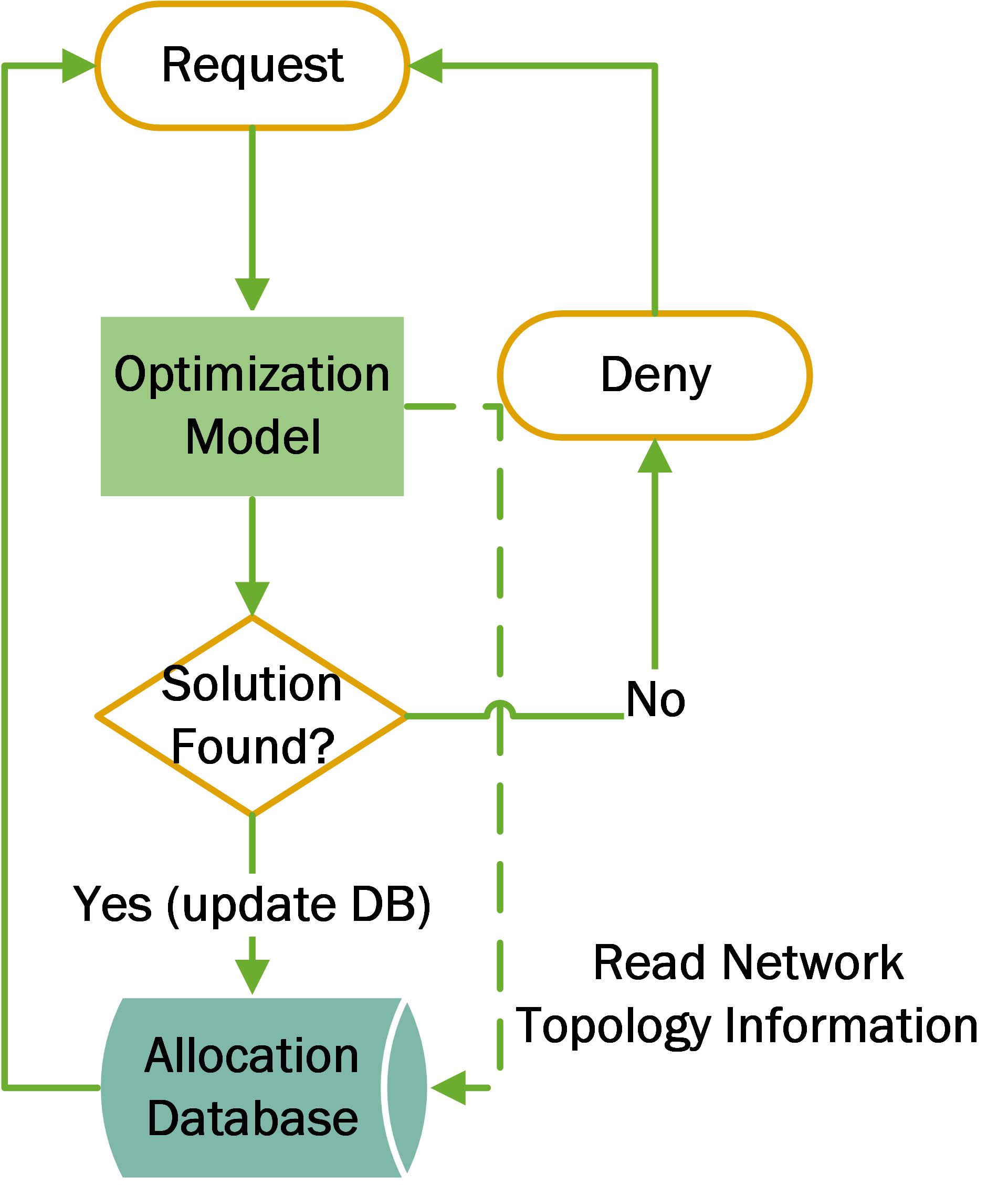}
	\caption{Dynamic Slice Allocation Optimization Flow Diagram}
	\label{fig:FrameworkAllocation}
\end{figure}

\section{DSAF: Dynamic Slice Allocation Framework}
\label{sec:DSAF:DSAF}
To automate the process of slice allocation in 5G core network, we propose a framework. In the proposed framework, slices can be allocated and deallocated dynamically. The \textbf{D}ynamic \textbf{S}lice \textbf{A}llocation \textbf{F}ramework (DSAF) consists of five components as shown in Fig.~\ref{fig:SDNTopology}.

\begin{itemize}
	\item \textbf{The Orchestrator:} The Orchestrator or the slice manager is responsible for managing slices and facilitating on-demand slice allocation, deallocation\footnote{DSAF has capability to deallocate slices. However, we have not performed experimented to show the deallocation of slices at the time of writing this paper} and coordinating different components of the framework as well as user interactions. 
	\item  \textbf{Optimization Module:} The optimization module implements our optimization model \cite{secureslicingCNS2019}. It reads the current state of the system allocation that includes remaining CPU, link bandwidth and delays as well as network topology and processes the incoming request. If a solution is found, the allocation is stored in a database, and the system allocation statistics are updated accordingly. 
	\item \textbf{Database (DB):} The database stores requests information, allocation schemes, renaming system resources, and performance statistics. 
	\item \textbf{O Agent:} The Optimization agent (O Agent) is responsible for communicating with the orchestrator and the optimization module. It receives the slice allocation from the orchestrator and forwards it to the optimization module and communicate back the results.
	\item \textbf{H Agent:} The Hypervisor Agent (H Agent) is an integral part of the framework (runs on each hypervisor). It is responsible for allocating and deallocating slices in real-time, starting applications for each slice and reporting slice statistics to the orchestrator.
\end{itemize}

The dynamic slice allocation process is shown in Fig. \ref{fig:SDNTopology} and it is described in the following steps (each step corresponds to the circled number in Fig. \ref{fig:SDNTopology}):
\begin{enumerate}
	\item The orchestrator provides user-interaction and waits for a slice request
	\item Once a slice request is received, it interacts with the O Agent to find the allocation scheme. 
	\item The O Agent starts an instance of the optimization model and pass the slice request to the optimization module. The optimization module reads the current network state from the database and finds the best solution to allocate the slice (if feasible). If no solution is found, the request is denied, and the response is sent to the orchestrator. The flow diagram of the optimization process is shown in Fig.~\ref{fig:FrameworkAllocation}
	\item If a solution is found in step 3, then the slice allocation will be stored in a database.
	\item The orchestrator receives an accepted or denied response from the O Agent. 
	\item If the response is slice request accepted, the orchestrator retrieves the allocation scheme from the database and sorts the retrieved allocation scheme according to the slice(s)  that will be allocated on each hypervisor. 
	\item The orchestrator sends the information to target H Agent(s). The information includes slice name, IP address, CPU (GHz), HDD, RAM, bandwidth and any application to start after the creation of VNF. 
	\item If the allocation is successful, the H Agent sends a successful response to the orchestrator
\end{enumerate}
\begin{figure}[!ht]
	\centering
	\includegraphics[width=\linewidth,keepaspectratio=true]{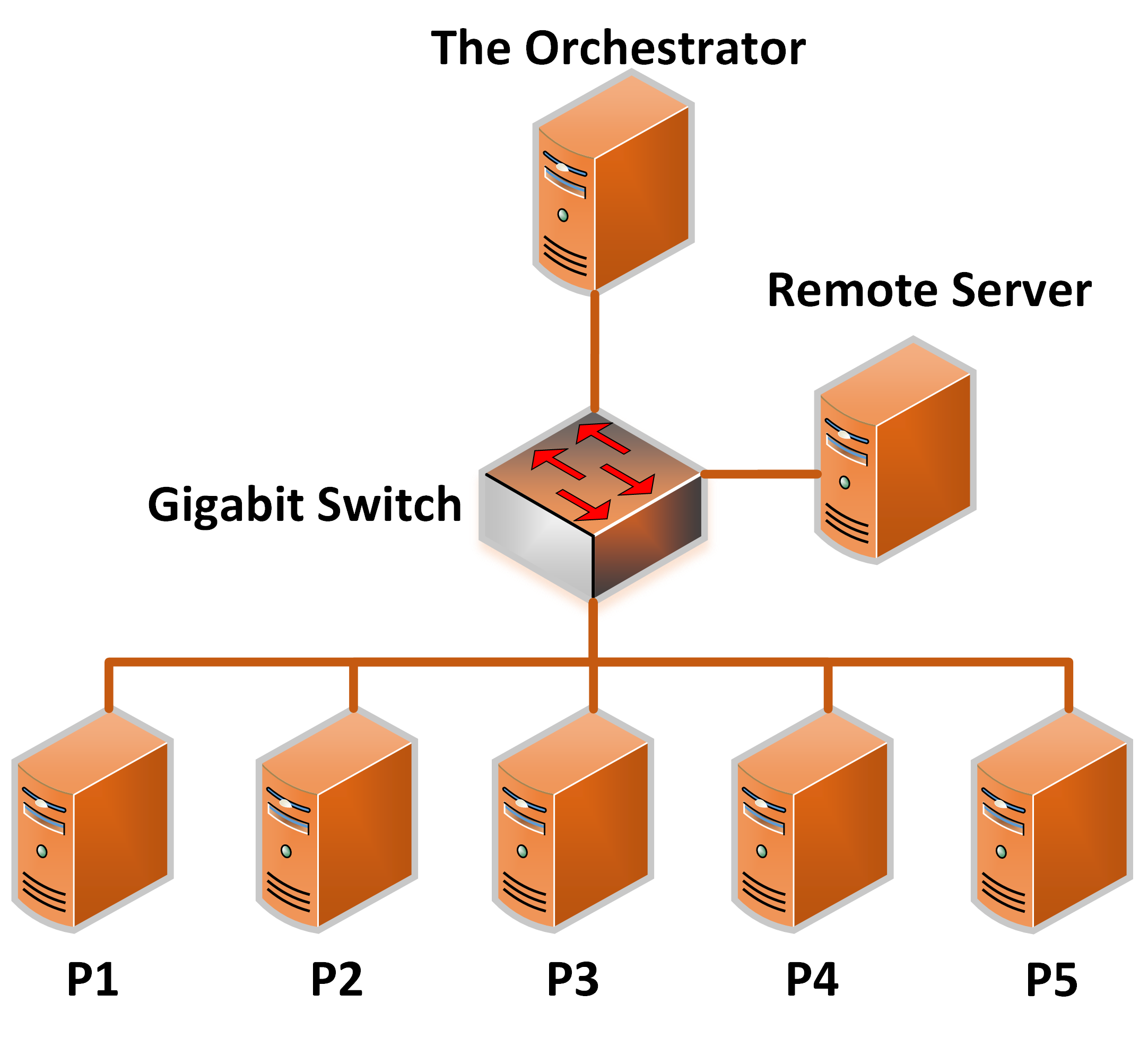}
	\caption{DSAF Physical Topology. All links are 1 Gbps }
	\label{fig:DSAF:physicalTopology}
\end{figure}
This process is repeated for every request, although DSAF should be able to process multiple requests at the same time.

\section{Experimental Setup}
\label{sec:DSAF:expsetup}

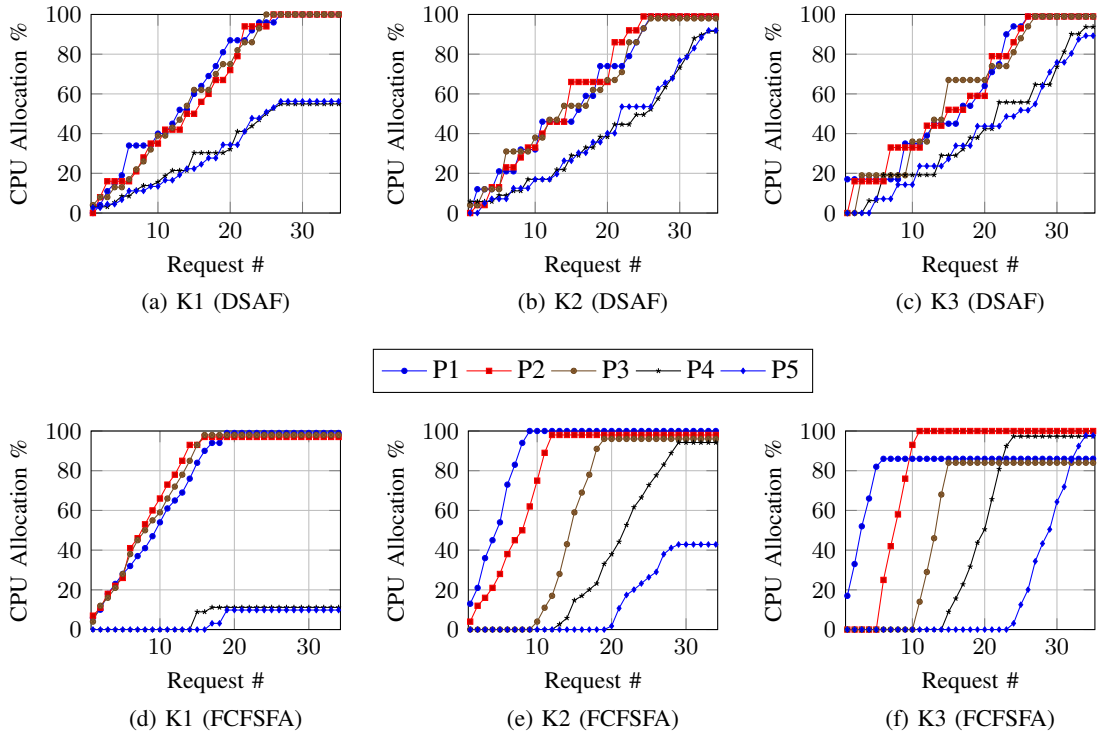
\begin{figure*}[!ht]
%	\caption{Comparison of Allocation Schemes for Different Scenarios}
%	\label{fig:comparitiveSetp}
	\begin{center}
		\begin{tikzpicture}
		
		\begin{groupplot}[
		group style={
			group name=my plots,
			group size=3 by 2,
			%			xlabels at=edge bottom,
			%ylabels at=edge left,
			horizontal sep=1.7cm,
			vertical sep=2.9cm,
		},
		xtick distance=10,
		ylabel = {CPU Allocation \%},
		xlabel={Request \#},
		grid=major,
		%x tick label style={rotate=45,anchor=east},
		ymax=100,
		ymin=0,
		width=0.27\linewidth
		]
		\nextgroupplot[legend to name={CommonLegend},legend style={legend columns=5}]
		xtick distance=10,
		\addplot+ [mark size=1pt] table[x=Index,y=P1] {K1.txt};
		\addplot+ [mark size=1pt] table[x=Index,y=P2] {K1.txt};
		\addplot+ [mark size=1pt] table[x=Index,y=P3] {K1.txt};
		\addplot+ [mark size=1pt] table[x=Index,y=P4] {K1.txt};
		\addplot+ [mark size=1pt] table[x=Index,y=P5] {K1.txt};
		\addlegendentry{P1}
		\addlegendentry{P2}
		\addlegendentry{P3}
		\addlegendentry{P4}
		\addlegendentry{P5}
		%	\captionof{subfigure}{abc}
		\nextgroupplot
		\addplot+ [mark size=1pt] table[x=Index,y=P1] {K2.txt};
		\addplot+ [mark size=1pt] table[x=Index,y=P2] {K2.txt};
		\addplot+ [mark size=1pt] table[x=Index,y=P3] {K2.txt};
		\addplot+ [mark size=1pt] table[x=Index,y=P4] {K2.txt};
		\addplot+ [mark size=1pt] table[x=Index,y=P5] {K2.txt};
		%Sistema para un kernel lineal:
		\nextgroupplot
		\addplot+ [mark size=1pt] table[x=Index,y=P1] {K3.txt};
		\addplot+ [mark size=1pt] table[x=Index,y=P2] {K3.txt};
		\addplot+ [mark size=1pt] table[x=Index,y=P3] {K3.txt};
		\addplot+ [mark size=1pt] table[x=Index,y=P4] {K3.txt};
		\addplot+ [mark size=1pt] table[x=Index,y=P5] {K3.txt};
		\nextgroupplot
		\addplot+ [mark size=1pt] table[x=Index,y=P1] {FCFSK1.txt};
		\addplot+ [mark size=1pt] table[x=Index,y=P2] {FCFSK1.txt};
		\addplot+ [mark size=1pt] table[x=Index,y=P3] {FCFSK1.txt};
		\addplot+ [mark size=1pt] table[x=Index,y=P4] {FCFSK1.txt};
		\addplot+ [mark size=1pt] table[x=Index,y=P5] {FCFSK1.txt};
		
		\nextgroupplot
		\addplot+ [mark size=1pt] table[x=Index,y=P1] {FCFSK2.txt};
		\addplot+ [mark size=1pt] table[x=Index,y=P2] {FCFSK2.txt};
		\addplot+ [mark size=1pt] table[x=Index,y=P3] {FCFSK2.txt};
		\addplot+ [mark size=1pt] table[x=Index,y=P4] {FCFSK2.txt};
		\addplot+ [mark size=1pt] table[x=Index,y=P5] {FCFSK2.txt};
		\nextgroupplot
		\addplot+ [mark size=1pt] table[x=Index,y=P1] {FCFSK3.txt};
		\addplot+ [mark size=1pt] table[x=Index,y=P2] {FCFSK3.txt};
		\addplot+ [mark size=1pt] table[x=Index,y=P3] {FCFSK3.txt};
		\addplot+ [mark size=1pt] table[x=Index,y=P4] {FCFSK3.txt};
		\addplot+ [mark size=1pt] table[x=Index,y=P5] {FCFSK3.txt};
		\end{groupplot}
		\path (my plots c1r1.south east) -- node[yshift=-21mm]{\ref{CommonLegend}} (my plots c3r1.south west);
		\node[text width=6cm,align=center,anchor=north] at ([yshift=-7mm]my plots c1r1.south) {\subcaption{K1 (DSAF)}\label{subplot:K1}};
		\node[text width=6cm,align=center,anchor=north] at ([yshift=-7mm]my plots c2r1.south) {\subcaption{K2 (DSAF)\label{subplot:K2}}};
		\node[text width=6cm,align=center,anchor=north] at ([yshift=-7mm]my plots c3r1.south) {\subcaption{K3 (DSAF)\label{subplot:K3}}};
		\node[text width=6cm,align=center,anchor=north] at ([yshift=-7mm]my plots c1r2.south) {\subcaption{K1 (FCFSFA)\label{subplot:FCFSK1}}};
		\node[text width=6cm,align=center,anchor=north] at ([yshift=-7mm]my plots c2r2.south) {\subcaption{K2 (FCFSFA)\label{subplot:FCFSK2}}};
		\node[text width=6cm,align=center,anchor=north] at ([yshift=-7mm]my plots c3r2.south) {\subcaption{K3 (FCFSFA)\label{subplot:FCFSK3}}};
		\end{tikzpicture} 
	\end{center}
	\caption{Comparison of Allocation Schemes for Different Scenarios}\label{fig:comparitiveSetp}
\end{figure*}

\begin{table}[!ht]
	\centering
	\caption{Experiment Topology Hardware Specification} \label{tab:SDNHardware}
	\begin{tabular}{c||c}\toprule[1.5pt]
		\rowcolor{Gray}
		\bf Server(s)         & \bf Hardware Specs     \\ \midrule
		
		Remote Server                 &  \makecell{CPU: 8 Cores,\\RAM: 32GB\\Bandwidth: 1Gbps}            \\\hline 
		\rowcolor{Gray}
		P1 to P3, and the orchestrator     &     \makecell{CPU: 4 cores,\\RAM: 8GB,\\Bandwidth: 1Gbps}    \\\hline
		P4 and P5     &     \makecell{CPU: 8 cores,\\RAM: 8GB, \\Bandwidth: 1Gbps}    \\\hline
		%\rowcolor{Gray}
		\bottomrule[1.25pt]
	\end{tabular}\par
	\bigskip
\end{table}

\begin{table}[!ht]
	\centering
	\caption{Total Available Resources for Allocation Across All Servers} \label{tab:AvailableResources}
	\begin{tabular}{c||c}\toprule[1.5pt]
		\rowcolor{Gray}
		\bf Resource        & \bf Capacity     \\ \midrule
		CPU Cores           &  28 Cores   \\ \hline 
		\rowcolor{Gray}
		CPU (GHz) 			& 74.8 GHz		\\\hline 
		RAM     			&     40 GB   \\\hline 
		\rowcolor{Gray}
		Hard Drive     		&     1TB   \\\hline 
		Total Bandwidth		& 5 Gbps \\\hline 
		\bottomrule[1.25pt]
	\end{tabular}\par
	\bigskip
\end{table}
To evaluate DSAF, we created a testbed using seven servers. We used five servers (P1 to P5) to allocate slices as shown in Fig. \ref{fig:DSAF:physicalTopology}. The optimization module and the database are hosted on the same server (Remote Server), and the orchestrator is hosted on a separate server. The hardware specification for all nodes are listed in Table~\ref{tab:SDNHardware} and the total resources available for the allocation across all servers are listed in Table~\ref{tab:AvailableResources}. The slice request parameters are listed in Table~\ref{tab:SliceRequestparamter}. For simplicity, each request arrives every three seconds (random interval can also be used) and they are allocated in the order of arrival. We note that slice requests do not expire.

\begin{table}[!ht]
	\centering
	\caption{Slice Request Parameters} \label{tab:SliceRequestparamter}
	\begin{tabular}{r||l}\toprule[1.5pt]
		
		\bf Parameter         & \bf Value     \\\midrule
		\rowcolor{Gray}
		Intra-slice isolation              &    1,2, or 3             \\\hline 
		VNF/Slice                 &    3              \\\hline 
		\rowcolor{Gray}
		Bandwidth            &  40-60 Mbps       \\\hline 
		CPU                &  0.75-2 GHz    \\\hline 
		\rowcolor{Gray}
		Total Slice Requests &     34         \\\hline 
		\bottomrule[1.25pt]
		\end {tabular}\par
		\bigskip
	\end{table}
We implemented the DSAF in Python~\cite{python2019} and MATLAB~\cite{MATLAB}. The orchestrator, O and H agents are written in Python, and the Optimization module is written MATLAB.  AMPL \cite{AMPL} is used to model the optimization algorithm, and CPLEX 12.9.0 is used as MILP solver.  OpenVZ \cite{OpenVZ} is used for virtualization. It is an open source container-based virtualization platform. OpenVZ allows each container to have a specific amount of CPU, RAM, and Hard Drive (HDD). Each container (which hosts one VNF) performs and executes like a stand-alone server. We have installed the CentOS 6 \cite{CentOS} operating system in every container. We used Linux Traffic control (\emph{tc}) \cite{tc} to allocate bandwidth for each container.
\section{Results and Discussion}
\label{sec:DSAF:resutls}

\begin{table}[!ht]
	\centering
	\caption{Experiment Scenarios} \label{tab:DSAF:scenarios}
	\begin{tabular}{c||l}\toprule[1.5pt]
		\rowcolor{Gray}
		\bf Scenario        & \bf Description     \\ \midrule
		K1          &  one VNF/hypervisor/slice   \\ 
		\rowcolor{Gray}
		K2 			& two or less VNFs/hypervisor/slice		\\
		K3     			&    three or less VNFs/hypervisor/slice   \\
		\rowcolor{Gray}
		\bottomrule[1.25pt]
	\end{tabular}\par
	\bigskip
	
\end{table}
We used three scenarios to allocate slices as listed in Table \ref{tab:DSAF:scenarios}. In each scenario, we compare the DSAF with the First Come First Serve First Available (FCFSFA).  In all scenarios, we collected statistics in terms of total slice requests allocated, processing time, and average computation time per slice. In the first scenario, we restricted the allocation to one VNF/hypervisor per slice ($K1$). In the second scenario, only two or fewer VNFs of a slice ($K2$) can be allocated on a single hypervisor. The third scenario three or less VNFs/slice can be placed on one hypervisor ($K3$). 
In FCFSFA, a slice is allocated based on arrival time and first available server. We make sure that allocated resources do not exceed the available physical resources. However, FCFSFA cannot guarantee the end-to-end delay. We wrote a Python script to perform allocation for FCFSFA. The FCFSFA is implemented on the same server as the orchestrator.
\pgfplotstableread[row sep=\\,col sep=&]{
	K    & DSAF & FCFS\\
	1   & 79.4 & 50\\
	2   & 100 & 85.3\\
	3	& 100 & 100\\
}\mydataSliceAllocated

\begin{figure}[ht!]
	\centering
	\begin{tikzpicture}[baseline]
	\begin{axis}[
	ybar,
	ymin=0,
	grid=major,
	ylabel={Requests Allocated (\%)},
	xlabel={Allocation Scenario},
	xticklabels={$K1$,$K2$,$K3$},
	legend style={at={(1,2)},
		anchor=north,legend columns=-1},
	legend style={font=\scriptsize,at={(0.5,-0.24)},
		anchor=north,legend columns=-1},
	xtick=data,
	%nodes near coords,
	]
	\addplot table[x=K,y=DSAF]{\mydataSliceAllocated};
	\addplot table[x=K,y=FCFS]{\mydataSliceAllocated};
	\legend{DSAF, FCFSFA}
	\end{axis}
	\end{tikzpicture}
	\caption{Total Slice Requests Allocated}
	\label{fig:RequestsAllocated}
\end{figure}
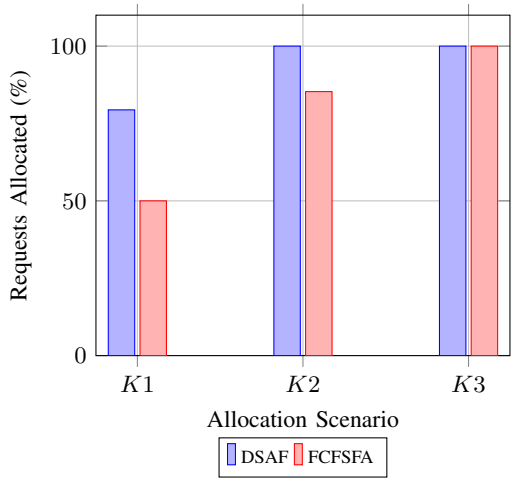

\pgfplotstableread[row sep=\\,col sep=&]{
	K    & DSAF & FCFSFA\\
	1   & 9.7751 & 9.143\\
	2   & 9.3737 &  9.157\\
	3   & 10.1368 & 9.159\\
}\mydataProcessingTime
\begin{figure}[ht!]
	\centering
	\begin{tikzpicture}[baseline]
	\begin{axis}[
	ybar,
	ymin=0,
	grid=major,
	ylabel={Time (millisec)},
	xlabel={Allocation Scenario},
	xticklabels={$K1$,$K2$,$K3$},
	legend style={at={(1,2)},
		anchor=north,legend columns=-1},
	legend style={font=\scriptsize,at={(0.5,-0.24)},
		anchor=north,legend columns=-1},
	%symbolic x coords={1,2,3,4,5,6,7,8,9,10},
	xtick=data,
	%nodes near coords,
	]
	\addplot table[x=K,y=DSAF]{\mydataProcessingTime};
	\addplot table[x=K,y=FCFSFA]{\mydataProcessingTime};
	\legend{DSAF, FCFSFA}
	\end{axis}
	\end{tikzpicture}
	\caption{Overhead: Processing Time}
	\label{fig:OverheadProcessing}
\end{figure}
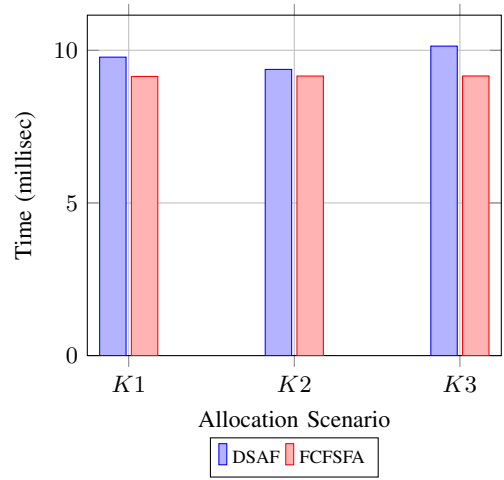
Fig~\ref{fig:comparitiveSetp} shows slice allocation for all scenarios. In scenario K3, all requests are successfully allocated for both allocation schemes as shown in Fig. \ref{subplot:K3} and \ref{subplot:FCFSK3}. However in scenario K1, DSAF and FCFSFA only allocated 27 and 17 slice requests respectively before P1, P2 and P3 ran out of CPU capacity. For scenario K1, we can only allocate one VNF/hypervisor per slice therefore once these three hypervisors ran out of CPU capacity we cannot allocate any more slices even though P4 and P5 still have significant resources available (because each slice needs three hypervisor for allocation). An interesting observation to note in Fig.~\ref{fig:comparitiveSetp} is that FCFSFA (all scenarios) allocates slices in an unbalanced manner. This allocation scheme behaves like a greedy approach, where it will allocate slices at the first available hypervisor. It resulted in lower number of requests being allocated in K1 and K2 as shown in Fig.~\ref{subplot:FCFSK1} and \ref{subplot:FCFSK2} respectively. It could also result in slices competing for resources on one hypervisor sooner even though the rest of the system is idle as well as a higher chance of slice unavailability if a hypervisor malfunctions. Whereas DSAF optimally allocates slices in all scenarios and spreads them across the entire system leading to less resource contention between slices and in case of a hypervisor malfunction, there is a higher chance that slices could remain partially or fully available. 

DSAF can allocate more or equal number of slice requests in all scenarios as shown in Fig.~\ref{fig:RequestsAllocated}.

Fig. \ref{fig:OverheadProcessing} shows the DSAF and FCFSFA processing time overhead. The processing time includes the time required to process the user requests, sending and receiving information from the H and O agents. For FCFSFA, the processing time is the time required to retrieve allocation requests and read system topology. Although DSAF requires slightly more processing time because of the communication required between the components of the framework, it still performs comparably to the FCFSFA in all scenarios. 
\pgfplotstableread[row sep=\\,col sep=&]{
	K    & DSAF & FCFSCA\\
	1   & 40 & 30\\
	2   & 70 & 35\\
	3   & 25 & 20\\
}\mydataAverageTime
\begin{figure}[ht!]
	\centering
	\begin{tikzpicture}[baseline]
	\begin{axis}[
	ybar,
	%ymode = log,
	ymin=0,
	grid=major,
	ylabel={Time (millisec)},
	xlabel={Allocation Scenario},
	legend style={at={(1,2)},
		anchor=north,legend columns=-1},
	legend style={font=\scriptsize,at={(0.5,-0.24)},
		anchor=north,legend columns=-1},
	%symbolic x coords={1,2,3,4,5,6,7,8,9,10},
	xtick=data,
	xticklabels={$K1$,$K2$,$K3$},
	%nodes near coords,
	]
	\addplot table[x=K,y=DSAF]{\mydataAverageTime};
	\addplot table[x=K,y=FCFSCA]{\mydataAverageTime};
	\legend{DSAF, FCFSCA}
	\end{axis}
	\end{tikzpicture}
	\caption{Overhead: Average Computation Time Per Slice}
	\label{fig:AveragecomputationTime}
\end{figure}
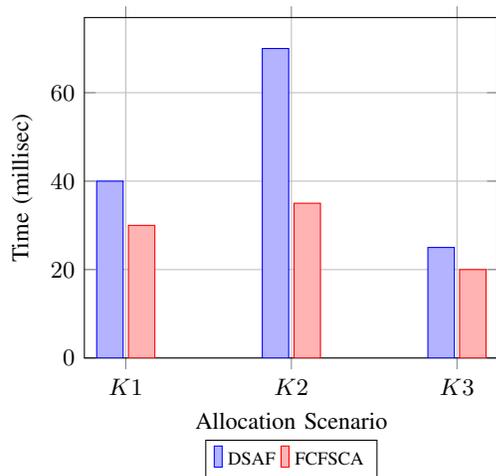

The average computation time per slice for DSAF is measured in the optimization module. The average computation time per slice for FCFSFA is the time required to calculate the allocation of slices and updating DB records. FCFSFA have lower average computation times per slice because there is no optimization performed as shown in Fig. \ref{fig:AveragecomputationTime}. DSAF's average computation time per slice is the cost of allocating more slices as well as providing flexibility when allocating slices.

\section{Conclusion}
\label{sec:DSAF:con}
In this paper, we presented a dynamic slice allocation framework to allocate slices in a resource efficient manner. The framework provides automation for slice allocation. We compared our framework with the First Come First Serve First Available allocation scheme. The evaluation of both techniques was done on a real testbed. Our results show that the overall proposed framework have comparable overhead to the FCFSFA. The cost of running DSAF is the average computation time that is slightly higher then FCFSFA. However, DSAF can allocate significantly more slices as well as it can fulfill a few requirements of the 5G (e.g., end-to-end delay). DSAF allocate slices in a balanced manner across the network which means less resource contention between slices until the network reaches saturation state. In FCFSFA, the resource contention could happen prematurely because slices are allocated in an unbalanced fashion (i.e., more slices on one hypervisor then the other).

\section*{Acknowledgment}
The second author acknowledges funding from Canada’s NSERC through the Discovery Grant Program.

\bibliographystyle{IEEEtran}
\bibliography{5Gbib}

\begin{thebibliography}{10}
\providecommand{\url}[1]{#1}
\csname url@samestyle\endcsname
\providecommand{\newblock}{\relax}
\providecommand{\bibinfo}[2]{#2}
\providecommand{\BIBentrySTDinterwordspacing}{\spaceskip=0pt\relax}
\providecommand{\BIBentryALTinterwordstretchfactor}{4}
\providecommand{\BIBentryALTinterwordspacing}{\spaceskip=\fontdimen2\font plus
\BIBentryALTinterwordstretchfactor\fontdimen3\font minus
  \fontdimen4\font\relax}
\providecommand{\BIBforeignlanguage}[2]{{%
\expandafter\ifx\csname l@#1\endcsname\relax
\typeout{** WARNING: IEEEtran.bst: No hyphenation pattern has been}%
\typeout{** loaded for the language `#1'. Using the pattern for}%
\typeout{** the default language instead.}%
\else
\language=\csname l@#1\endcsname
\fi
#2}}
\providecommand{\BIBdecl}{\relax}
\BIBdecl

\bibitem{8116371}
R.~Ford, A.~Sridharan, R.~Margolies, R.~Jana, and S.~Rangan, ``{Provisioning
  low latency, resilient mobile edge clouds for 5G},'' in \emph{2017 IEEE
  Conference on Computer Communications Workshops (INFOCOM WKSHPS)}, May 2017,
  pp. 169--174.

\bibitem{ICT-31766}
\BIBentryALTinterwordspacing
{P. Popovski and V. Brau and H.-P. Mayer and P. Fertl and Z. Ren and D.
  Gonzales-Serrano and E. G. Strom and T. Svensson and H. Taoka and P.
  Agyapong}, ``{EU FP7 INFSO-ICT-317669 METIS, D1.1: Scenarios, requirements
  and KPIs for 5G mobile and wireless system},'' {Mobile and wireless
  communications Enablers for the Twenty-twenty Information Society}, TS~{1},
  April 2013. [Online]. Available:
  \url{http://publications.lib.chalmers.se/records/fulltext/213055/local\_213055.pdf}
\BIBentrySTDinterwordspacing

\bibitem{rfcns}
\BIBentryALTinterwordspacing
{X. de Foy and A. Rahman}, ``{Network Slicing - 3GPP Use Case},'' Internet
  Requests for Comments, {RFC Editor}, {RFC}, October 2017. [Online].
  Available:
  \url{https://tools.ietf.org/id/draft-defoy-netslices-3gpp-network-slicing-02.html}
\BIBentrySTDinterwordspacing

\bibitem{8676260}
D.~{Sattar} and A.~{Matrawy}, ``{Optimal Slice Allocation in 5G Core
  Networks},'' \emph{{IEEE Networking Letters (Early Access)}}, 2019.

\bibitem{secureslicingCNS2019}
D.~Sattar and A.~Matrawy, ``{Towards Secure Slicing: Using Slice Isolation to
  Mitigate DDoS Attacks on 5G Core Network Slices},'' to appear in {7th Annual
  IEEE Conference on Communications and Network Security (CNS 2019)}, June
  2019.

\bibitem{7499297}
M.~{Jiang}, M.~{Condoluci}, and T.~{Mahmoodi}, ``Network slicing management
  amp; prioritization in 5g mobile systems,'' in \emph{European Wireless 2016;
  22th European Wireless Conference}, May 2016, pp. 1--6.

\bibitem{7509393}
X.~Zhou, R.~Li, T.~Chen, and H.~Zhang, ``Network slicing as a service: enabling
  enterprises' own software-defined cellular networks,'' \emph{IEEE
  Communications Magazine}, vol.~54, no.~7, pp. 146--153, July 2016.

\bibitem{7996490}
M.~Jiang, M.~Condoluci, and T.~Mahmoodi, ``{Network slicing in 5G: An
  auction-based model},'' in \emph{2017 IEEE International Conference on
  Communications (ICC)}, May 2017, pp. 1--6.

\bibitem{Nikaein:2015:NSE:2795381.2795390}
N.~Nikaein, E.~Schiller, R.~Favraud, K.~Katsalis, D.~Stavropoulos, I.~Alyafawi,
  Z.~Zhao, T.~Braun, and T.~Korakis, ``Network store: Exploring slicing in
  future 5g networks,'' in \emph{Proceedings of the 10th International Workshop
  on Mobility in the Evolving Internet Architecture}, ser. MobiArch '15, 2015,
  pp. 8--13.

\bibitem{7116162}
A.~Baumgartner, V.~S. Reddy, and T.~Bauschert, ``{Mobile core network
  virtualization: A model for combined virtual core network function placement
  and topology optimization},'' in \emph{Proceedings of the 2015 1st IEEE
  Conference on Network Softwarization (NetSoft)}.\hskip 1em plus 0.5em minus
  0.4em\relax {IEEE}, April 2015, pp. 1--9.

\bibitem{8524891}
G.~{Garcia-Aviles}, M.~{Gramaglia}, P.~{Serrano}, and A.~{Banchs}, ``Posens: A
  practical open source solution for end-to-end network slicing,'' \emph{IEEE
  Wireless Communications}, vol.~25, no.~5, pp. 30--37, October 2018.

\bibitem{8491249}
L.~Zanzi and V.~Sciancalepore, ``{On Guaranteeing End-to-End Network Slice
  Latency Constraints in 5G Networks},'' in \emph{{15th International Symposium
  on Wireless Communication Systems (ISWCS)}}, 2018.

\bibitem{python2019}
\BIBentryALTinterwordspacing
{Python Software Foundation}, ``Python,'' accessed 27 April 2019. [Online].
  Available: \url{{https://www.python.org/}}
\BIBentrySTDinterwordspacing

\bibitem{MATLAB}
\BIBentryALTinterwordspacing
MathWorks, ``{MATLAB},'' {April 2019}. [Online]. Available:
  \url{https://www.mathworks.com/products/matlab.html}
\BIBentrySTDinterwordspacing

\bibitem{AMPL}
\BIBentryALTinterwordspacing
``{A Modeling Language for Mathematical Programming},'' {January 2019}.
  [Online]. Available: \url{https://ampl.com/}
\BIBentrySTDinterwordspacing

\bibitem{OpenVZ}
\BIBentryALTinterwordspacing
``{OpenVZ},'' {January 2019}. [Online]. Available: \url{https://openvz.org/}
\BIBentrySTDinterwordspacing

\bibitem{CentOS}
\BIBentryALTinterwordspacing
``{CentOS},'' {January 2019}. [Online]. Available:
  \url{https://www.centos.org/}
\BIBentrySTDinterwordspacing

\bibitem{tc}
\BIBentryALTinterwordspacing
B.~Hubert, ``{Traffic Control (tc)},'' {January 2019}. [Online]. Available:
  \url{https://linux.die.net/man/8/tc}
\BIBentrySTDinterwordspacing

\end{thebibliography}
\end{document}